# Molecular fingerprinting of biological nanoparticles with a label-free optofluidic platform

*Alexia Stollmann,[1] Jose Garcia-Guirado,[1] Jae-Sang Hong,[2] Hyungsoon Im,[2,3] Hakho Lee,[2,3] Jaime Ortega Arroyo,[1]\* Romain Quidant[1]\**

[1] Nanophotonic Systems Laboratory, Department of Mechanical and Process Engineering, ETH Zurich, 8092 Zurich, Switzerland.

[2] Center for Systems Biology, Massachusetts General Hospital, Boston, Massachusetts 02114, United States.

[3] Department of Radiology, Massachusetts General Hospital, Boston, Massachusetts 02114, United States.

*Corresponding authors: jarroyo@ethz.ch, rquidant@ethz.ch

**Abstract (250 words, no claims of novelty)**

Label-free detecting multiple analytes in a high-throughput fashion has been one of the long-sought goals in biosensing applications. Yet, for all-optical approaches, interfacing state-of-the-art label-free techniques with microfluidics tools that can process small volumes of sample with high throughput, and with surface chemistry that grants analyte specificity, poses a critical challenge to date. Here, we introduce an optofluidic platform that brings together state-of-the-art digital holography with PDMS microfluidics by using supported lipid bilayers as a surface chemistry building block to integrate both technologies. Specifically, this platform fingerprints heterogeneous biological nanoparticle populations via a multiplexed label-free immunoaffinity assay with single particle sensitivity. Herein, we first thoroughly characterise the robustness and performance of the platform, and then apply it to profile four distinct ovarian cell-derived extracellular vesicle populations over a panel of surface protein biomarkers, thus developing a unique biomarker fingerprint for each cell line. We foresee that our approach will find many applications where routine and multiplexed characterisation of biological nanoparticles is required.

**Keywords (5-7):** Extracellular vesicles, label-free imaging, supported lipid bilayer, immunoassays, microfluidics, multiplexing, holography

**Introduction.**
Accurate reconstruction of heterogeneous biological nanoparticle populations demands methods that satisfy three key parameters: sensitivity, high-throughput, and molecular fingerprinting. Extracellular vesicles (EVs), membrane-bound particles secreted by cells of all kinds[1,2], are a prime example of nanoparticle systems that would greatly benefit from a new generation of characterisation methods that simultaneously comply with these three requirements. This is because the smaller the size of a particle, the greater the demand on sensitivity, which usually is paid in the currency of throughput. Similarly, the greater the number of biomarkers to screen, the lower the throughput. Thus, the ideal approach in terms of sensitivity would be one which can detect these biological nanoparticles at the single particle level regardless of size in aqueous environments. Regarding the throughput, the approach should enable statistically significant sampling (> 10,000 events) within a reasonable time, i.e., on the time scale of minutes to an hour. Lastly, in terms of molecular fingerprinting, the approach should differentiate between subpopulations expressing relevant biomarkers and minimise the rate of false positive. To date, fluorescent-based single-particle assays are the most established and prevalent due to the intrinsic specificity and single-molecule sensitivity attained by fluorescence labelling and its compatibility with microfluidics, which capitalises on high throughput and minimal sample processing. Fluorescence-based molecular fingerprinting has so far been achieved through either sequential read-out of different fluorescent probes[3,4], spectral emission decoding[5], spatial patterning[6], or a combination thereof[7]. Despite widespread use, fluorescence-based detection has intrinsic limitations either in the form of labelling efficiency, fixed photon budget or labelling incompatibility[8]. As a result, there is a need for all-optical label-free alternatives compatible with high throughput microfluidics that can deliver all the benefits of single-molecule fluorescence assays without the constraints associated with labelling.

From the available all-optical label-free methods, those based on elastic scattering have become one of the most promising, as they now routinely achieve detection sensitivities down to the single protein[9–11], nucleic acid[12], and micelle level[13] that rival single-molecule fluorescence. Yet their translation to routine particle characterisation faces challenges in the form of throughput, specificity, and ability to perform multiplexed read-out. To understand this, it suffices to consider that the amount of light scattered from said biological nanoparticles pales in comparison to light scattered by the substrate roughness; thus, the extreme sensitivity of all these surface-based techniques hinges on an imaging modality whereby the static background from the observation area is constantly updated. Such imaging modality, termed in some cases as differential imaging, comes with a main drawback. namely, that only one region of the sensor can be observed during an assay. In addition, from an optics perspective, these approaches suffer from relatively small fields of view (FOV), which rarely exceed the scale of 100s of μm$^2$. These restricted sensing areas aim to minimise deleterious effects from either parasitic background scattering from the imaging optics, or unwanted interferences due to the coherent nature of the light source typically used. For systems that do not demand the highest sensitivity, i.e., in the absence of differential imaging, the scattering signal from the substrate roughness as well as unwanted interferences that may arise from multiple interfaces, a common scenario in microfluidic devices, set the lower limit of detection and should be minimised throughout any surface functionalisation step. Although the substrate roughness can be significantly reduced using atomically flat substrates like mica[14], this comes at the expense of losing target specificity. Conversely, introducing specificity to the surface in the form of a capture probe via surface chemistry, e.g., with an immobilised antibody or aptamer, increases the substrate roughness, and thereby restricts the detection sensitivity. Even in cases where surface chemistry has allowed multiplexed detection in a non-differential imaging mode, the sample had to be imaged in the air to enhance the scattering contrast relative to measuring in aqueous solutions[15]. Thus, the challenges remain to simultaneously deliver sensitivity, throughput, and specificity in aqueous solution-based label-free detection assays.

Microfluidic integration can address the throughput and multiplexing challenges associated with label-free detection approaches. Nevertheless, finding a functionalisation scheme that delivers target specificity and minimises non-specific binding without introducing additional unwanted scattering signals remains an important obstacle. Namely, despite the availability of numerous strategies, the straightforward assembly of PDMS-based microfluidic involves a step that exposes the functionalised substrate to harsh conditions, such as plasma treatment followed by baking at high temperature, which inevitably compromises the integrity of any functionalisation. A recent alternative has been demonstrated by using masks during the assembly process; however, the need for µm-level alignment between chip and the protective element, in the case of complex chip designs, imposes a steep technological restriction[16]. As a result, these state-of-the-art functionalisation approaches are incompatible with complex PDMS microfluidics, and in-situ/on-chip solutions should be sought. Nevertheless, existing in-situ alternatives require either long incubation periods on the time-scale of hours, as is the case of poly(ethylene)glycol (PEG)-based strategies; or compromise on the degree of passivation, for instance, bovine serum albumin[17].

Bringing together state-of-the-art all-optical label-free approaches with microfluidics requires an integrated solution that addresses the limitations intrinsic to each tool. In this work we present a label-free optofluidic platform that delivers a solution satisfying these three key parameters: sensitivity, high throughput, and molecular fingerprinting. Specifically, we first identified a surface functionalisation protocol, in the form of high-quality supported lipid bilayers (SLBs), that acts as a building block to integrate microfluidic technology with label-free detection with single-particle sensitivity. Using this building block we implemented a label-free immunoaffinity pull-down assay and assessed the performance of each stage of the functionalisation by taking advantage of the highly sensitive and label-free detection scheme of the platform. Finally, we showcase all the features of the platform by profiling populations of EVs from four different ovarian cell lines with single EV sensitivity in a multiplexed and label-free manner and thus generate characteristic fingerprints for each EV subpopulation based on a panel of surface biomarkers.

**Results and Discussion**
**Concept and experimental workflow**
In this work, we fulfilled the requirements for label-free molecular fingerprinting of heterogeneous nanoparticle suspensions by focussing our efforts around three main concepts: i) large FOV imaging with single particle sensitivity, ii) high throughput, small volume, and individually addressable microfluidic channels, and iii) an in-chip surface functionalisation protocol for pull-down immunoaffinity assays (Fig. 1).

For large FOV imaging, we used an inline holographic microscope in reflection geometry with an intrinsic requirement of a spatially incoherent light source as we are only interested in interferometric contributions between the surface and nanoparticles immobilised to it. Fig. 1B schematically depicts the optical read-out strategy. As an imaging area we targeted illumination FOVs on the order 100×100 µm$^2$, which are rarely achieved with iSCAT microscopy with high numerical aperture (NA) objectives due to the presence of detrimental parasitic fringes that arise from the reflections from multiple closely spaced interfaces in microfluidic chips. In addition to reducing these parasitic interferences when imaging through microfluidic chips[18], the spatially incoherent illumination drastically also drastically reduces the influence of speckles. To do so, the output from a narrowband fibre-coupled light emitting diode (LED) was relay imaged onto the sample, providing a total illumination area of 100×100 µm$^2$. Light scattered by the sample as well as the weak reflection from the substrate interface was collected by the high NA objective, and subsequently their interference imaged onto a camera. Such illumination scheme is not limited to LEDs as the similar performance was also obtained by reducing the spatial coherence of a diode laser with a combination of a rotating ground glass diffuser and a multimode fibre (Fig. S1). To extend the FOV we followed established computer vision routines to stitch a series of raster scanned images.

To satisfy the low volume reagent, multiplexing, and throughput requirement, we used PDMS microfluidic technology based on Quake microvalves[19] as shown in Fig. 1C. These chips were composed of a control (orange) and flow layer (light blue) to independently address different sensing channels (black arrows), and finely control each step of the immunocapture assay without interference from the user. Here each channel represented a different experiment programmatically controlled via a computer interface, thereby opening the possibility for long-term automation. To maintain uniform flow conditions all channels were designed with the same microfluidic resistance by keeping the dimensions of each channel fixed. Uniform flow rates are critical to guaranteeing consistent advection-driven kinetic conditions and minimising mass-transport limited effects throughout all the assays within a chip. In terms of total volume, each sensing area corresponded to 10 nL (length, width, height: 3 mm, 0.3 mm, 0.01 mm), which upon including the inlet and outlet path lengths, increased to approximately 40 nL per channel. Added together the whole microfluidic device operated with less than 0.5 µL of sample.

Surface chemistry was central to bringing both established microfluidics and imaging technologies under a common umbrella. In this work as shown in Fig. 1D, we opted for SLBs as the basis for the in-chip functionalisation protocol due to their biomimetic nature, ease of preparation, intrinsic anti-fouling properties, and on-chip compatibility[20–22]. The SLBs prepared by fusogenic-assisted vesicle fusion simultaneously acted as a passivating coating against non-specific binding, and as a building block for the immunocapture pull-down assay. Regarding the pull-down functionalisation scheme, NeutrAvidin molecules coupled biotinylated monoclonal antibodies to the SLBs doped with biotinylated lipids.

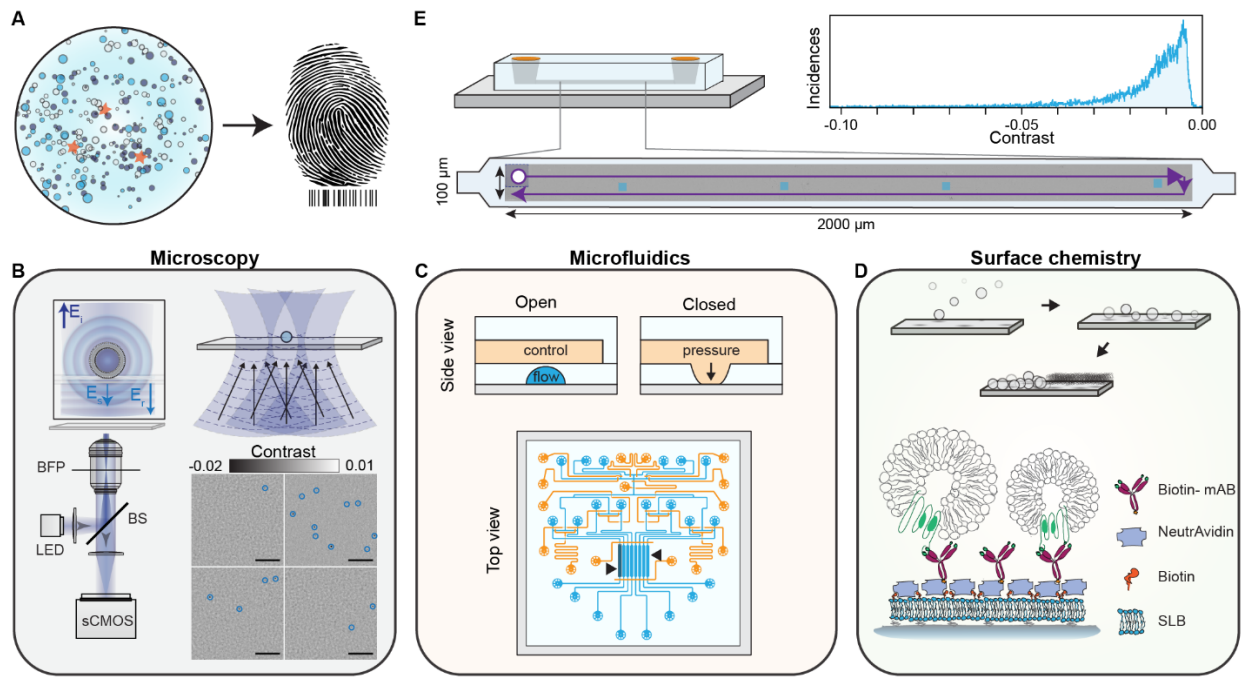

**Fig. 1. Concept and workflow of the label-free optofluidic platform. (A)** Conceptual illustration of the aim of the platform. The platform is based on three main toolboxes. **(B)** Microscopy toolbox: schematic of the optical system for large FOV imaging with single particle sensitivity based on spatially incoherent inline holography in a reflection geometry together with four representative zoomed-in images with diffraction limited spots identified with blue circles. Inset: the working principle relies on detecting the interference between the weakly scattered light from the sample, $E_s$, and the reflection from the substrate/water interface, $E_r$. Scale bars: 5 μm. **(C)** Microfluidic toolbox: representative two-layer microfluidic chip design composed of a network of valves (orange) and flow channels (blue). The black arrows highlight the section of independently addressable channels used for sensing. **(D)** Surface chemistry toolbox: schematic representation of the in-chip functionalisation scheme based on SLB formation by liposome fusion, which acts as the building block for the immunoaffinity pull-down assays. **(E)** Workflow of the platform: representative experimental image scan of a sensing channel obtained by stitching multiple fields-of-view together with the resulting contrast distribution of all localised single particles. The scattering contrast signals are retrieved upon localising all the diffraction-limited spots above a signal-to-noise ratio (SNR) threshold, an example shown in (B).

As a general workflow to either evaluate the performance at each stage of the functionalisation process or molecularly fingerprint EV populations, we raster scanned the sample along an area covering approximately 66% of the microfluidic sensing channel length (3 mm long) and stitched the acquired images together to generate image scans such as Fig. 1E. After flat field correction, diffraction-limited spots (corresponding to biological nanoparticles, surface inhomogeneities or defects) were localised, their signal contrast obtained and subsequently plotted to determine their contrast distribution (Fig. 1E). These large image scans allowed us to build robust statistics, increase throughput and identify inhomogeneities in the surface functionalisation protocol.

**Supported lipid bilayers (SLBs) as the building block for immunoaffinity assays**

The quality of the sensing substrate is of utmost importance for any label-free assays as unwanted scattering from imperfections or defects will contribute to a false positive readout. This is particularly critical in the case of SLBs if one considers the potential overlap in sizes between EVs and any remaining unruptured liposomes from the SLB formation[23–25]. Even in the case of sensing based on differential imaging[9,11,26,27], the presence of considerable scattering signals, such as large unruptured liposomes, impose tighter experimental constraints in the form of better sample stabilisation to compensate for the minute sample drifts that push the differential imaging approach away from the shot noise limited detection. To determine the most suitable lipid coating strategy, i.e., one that effectively reduces the likelihood of false positives during a sensing assay with high reproducibility, we screened different SLB preparation methods. As a metric we aimed to minimise the number of scattering signals present in the formed bilayer.

For the SLB formation, we chose the fusogenic agent-assisted bilayer formation strategies, as they are the most promising in generating high-quality continuous lipid coatings with minimal defects irrespective of lipid composition and substrate properties[28,29]. We specifically used the α-helical (AH) peptide as the fusogenic agent since buffer washes can fully remove it from the formed bilayer, and therefore not influence further downstream steps[28]. The fusogenic activity of the AH peptide depends on the membrane curvature of the unruptured liposomes, and thereby their size; with smaller liposomes having higher curvature and peptide activity[30–32]. To determine the liposome size distribution that leads to the most reproducible and suitable bilayer for label-free sensing, we tested different preparations based on either extrusion or bath sonication. We monitored the formation of the SLBs with emphasis on three key stages: the bare substrate in the presence of buffer solution (PBS), the initial bilayer formed after liposome fusion (liposome), and the final bilayer after AH peptide incubation and subsequent buffer rinsing (peptide) (Fig. 2A). One of the key advantages of this functionalisation scheme, when combined with microfluidics, is the speed of preparation, which occurs on the timescale of minutes (Movies S1-S2).

For extrusion, the polycarbonate membrane pore size tuned the liposome size from 30 to 200 nm; whereas, bath-sonication offered a minimal sample preparation at the expense of no control over the size distribution. Their respective size distributions determined by dynamic light scattering are shown in Fig. 2B, with liposomes prepared via extrusion displaying higher uniformity and reproducibility compared to bath sonicated ones. Fig. 2C shows representative zoom-in images, corresponding to an area of $20 \times 20~\mu m^2$ for each of the stages, to highlight the differences in SLB formation driven by substrate-vesicle interactions, as well as, between peptide-induced bilayer repair. The first row, corresponding to the buffer only step, provides an initial quality assessment of the cleaned glass substrate. At this stage of the process, we observed the presence of substrate roughness together with the inherent inhomogeneity of the substrate, either in the form of defects or contaminants. The second row shows the representative examples of the formed SLB after vesicle fusion and buffer rinsing to remove excess liposomes. Here, the effect of substrate-liposome interaction is most noticeable in the number and signal contrast of the diffraction limited spots. These diffraction limited spots were assigned as either membrane defects in the form of unruptured liposomes, trapped liposomes, and inhomogeneities in the bilayer, or defects already present in the bare substrate. Qualitatively, SLBs formed via liposome fusion alone favour larger liposome preparations (200 nm and sonicated), as they are more likely to rupture spontaneously compared to smaller ones[33]. The third and final row shows the bilayer after continuously flowing in the fusogenic peptide followed by an osmotic shock upon buffer exchange. In all cases, the bilayers treated with the AH peptide significantly reduced the number of membrane defects compared to those formed by liposome fusion alone.

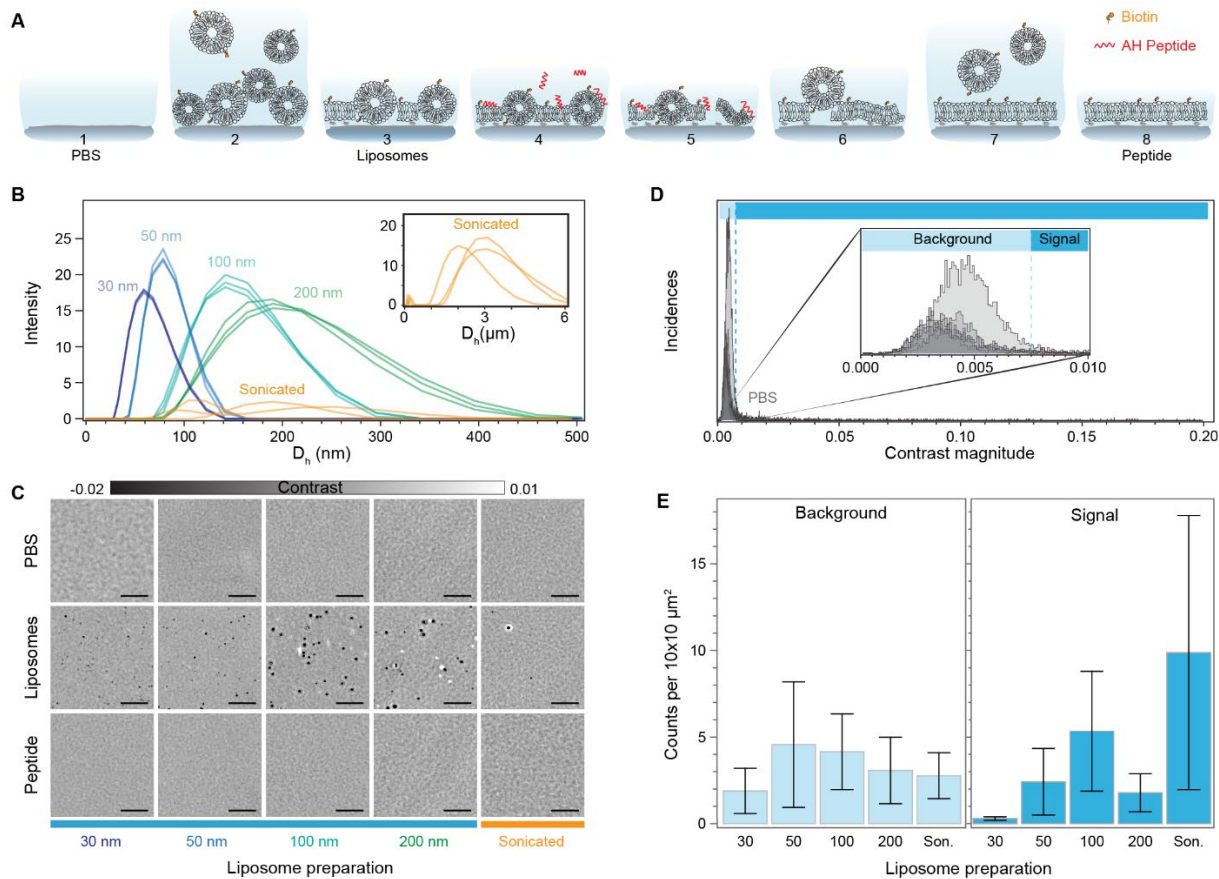

**Fig. 2. Characterisation of the number of bilayer defects. (A)** Diagram showing the steps involved in preparation of the supported lipid bilayer via fusogenic AH peptide interaction and osmotic stress. **(B)** Hydrodynamic size of the different liposome preparations as determined by dynamic light scattering. **(C)** Zoom-in of representative images for each preparation method at the different stages of the peptide-mediated supported lipid bilayer formation process. PBS: clean substrate exposed to only buffer solution; Liposomes: substrate after vesicle fusion and buffer rinsing; Peptide: supported lipid bilayer after peptide incubation and osmotic shock buffer rinsing. **(D)** Particle contrast histograms from all localisations found in substrates exposed to only a buffer solution (PBS). Each line corresponds to an approximate scanned area of 0.2 mm$^2$. **(E)** Particle localisation density as a function of SLB preparation and categorised according to the contrast falling within background and signal regions respectively. Each bar corresponds to the mean over N = (9,5,4,5,5) different substrates. Error bars represent the standard deviation over the mean. Scale bars: 5 μm

To assess the quality of the final bilayer, we determined the number of defects before and after bilayer formation within each sensing channel, and reported them in the form of density, i.e., counts per area of 10×10 μm$^2$. We chose this area to allow meaningful comparison amongst most state-of-the-art label-free detection schemes, which have FOVs with dimensions ranging in the tens of microns[27,34,35]. To do so, we performed image scans covering an area of approximately 0.2

mm$^2$ over a minimum of three different substrates for each liposome preparation. Here we assign defects to any diffraction limited signal that is 4× the noise floor. This SNR cut-off was selected to minimise the occurrence of false positives attributed to noise fluctuations. As a first step, we determined the baseline contrast distribution of defects present in the bare substrate with buffer. Fig. 2D shows that most defects fall within a narrow contrast range, with a cut-off contrast value of $7.5\times10^{-3}$ as indicated by the dashed vertical line. We attributed these to surface roughness and minute substrate inhomogeneities. We classified this narrow contrast region as background, and the one above said contrast threshold as the signal. Such classification minimises the influence of substrate roughness and small inhomogeneities in the defect density metric without the need for further image processing, i.e., background subtraction or differential imaging.

Across the different SLB preparations, we observed similar defect occurrences in the background signal region (Fig. 2E). However, in the detection sensing window, liposomes prepared via extrusion with a 30 nm polycarbonate pore size formed the most reproducible SLBs with also the smallest defect density. Although the other approaches led to potentially high-quality lipid coatings as shown in Fig. 2C, there was a high level of variability when assessed over larger observation areas, stressing once again the importance of large FOV imaging. We can rationalise these results by considering the membrane curvature sensitivity of the AH-peptide, which preferentially ruptures liposomes with diameters below 125-150 nm[32]. The size distribution from Fig. 2B confirms that the 30 nm liposome preparation has the smallest fraction of liposomes above this cut-off. As a corollary, the probability of having a higher proportion of unruptured liposomes unaffected by the AH peptide is significantly higher for all other preparations, thereby leading to a greater variance.

To put our results in the context of other works, our lowest defect density per 100 μm$^2$ is on the order 0.5 compared to the 0.05 counts previously reported[29]. The higher sensitivity and the label-free nature of our imaging platform can account for this discrepancy by considering analysing much larger areas makes it more statistically likely to find substrate defects and inhomogeneities.

**Microfluidic chip reproducibility and robustness**
Inter- and intra- chip reproducibility are critical for microfluidic-based assays. We assessed the reproducibility of the SLB formation process over multiple microfluidic chips (including different designs), and across different portions of the substrate. The microfluidic chips were designed to have between 4 to 8 independent flow channels, from which an imaging area equivalent to 0.2 mm$^2$ was scanned and the corresponding defect density determined. Fig. 3 shows that across five independent chips measured over different days after liposome preparation, the defect density in the signal region is highly reproducible both within and across microfluidic chips. Moreover, the presence of high background signal levels from the bare glass substrate, e.g., chip 3, did not affect this high degree of reproducibility in the signal area. Also, despite the number of defects in the signal region largely correlated with the underlying quality of the bare substrate, the average remained near the value reported in Fig. 2E.

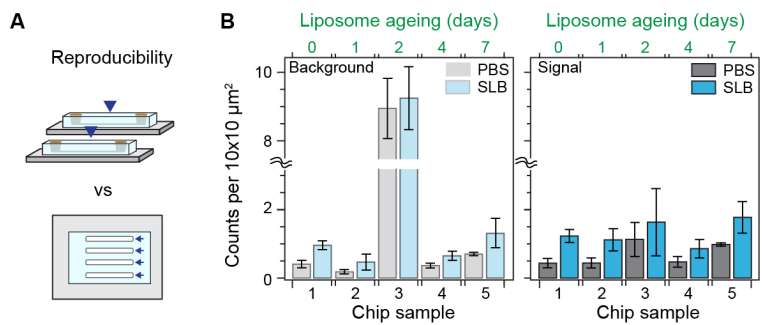

**Fig. 3. Robustness of the supported lipid bilayer. (A)** Cartoon depicting the difference between chip-to-chip (inter-) and within-chip (intra-) variability. **(B)** Number of defects within the expected background and signal contrast regions for different chips, before and after SLB formation. The number of days after liposome preparation for each chip appears on top. Each bar corresponds to the mean of scans of an approximate area of 0.2 mm$^2$ over multiple different channels within each chip (N > 4). Error bars represent the standard deviation over the mean.

As the liposome sample aged, we observed a decrease in the spontaneous rupture frequency alongside an increase in unruptured liposomes prior to peptide treatment (Fig. S2). Nonetheless, the defect density did not significantly change upon peptide treatment. These results highlight the robustness and flexibility of the platform to reagent ageing. This feature allows the decoupling of the liposome preparation steps from the bilayer formation ones; a critical aspect when dealing with microfluidic devices.

**Compatibility with on-chip immunoaffinity capture strategies**
To access the standard immunocapture functionalisation scheme based on NeutrAvidin as a linker between biotinylated antibody and lipid, all bilayers were composed of POPC : biotin DOPE lipids in a 99:1 molar ratio. At this molar ratio, we expected an almost complete antibody coverage of the substrate, given an estimated surface density of 1.4 biotins per 10×10 nm$^2$ (2.3 pmol/cm$^2$), slightly below the minimum doping to achieve a full monolayer of NeutrAvidin previously reported to occur at 2.8 biotins per 10×10 nm$^2$ (3.5% molar biotin, 8 pmol/cm$^2$)[36]. We chose POPC as the main phospholipid component in our liposome preparation based on its favourable physical properties for SLB formation, namely: zwitterionic nature, low melting temperature, preference to form lamellar rather than hexagonal structures, and cylindrical shape with little to no curvature[37–39].

To evaluate whether progressive functionalisation steps, i.e., the addition of NeutrAvidin followed by biotinylated antibodies, impact the bilayer quality, we quantified the number of defects accumulated at each step (Fig. 4A). In the contrast region assigned to the background, we observed no major differences in the quality of the bilayer other than a small rise in the number of defects, which we associated to an increase in surface roughness caused by the respective randomly oriented protein coatings (Fig. 4B). The signal region exhibited a slight increase in localisations after antibody incubation, yet without statistical significance (one-way ANOVA: Chip 1, $P = 0.109$; Chip 2, $P = 0.303$), likely attributed to aggregates (Fig. 4B).

We then validated the immunoassay by first flowing a sample of streptavidin-labelled 20 nm gold nanoparticles (AuNPs-SAv) and biotinylated-liposomes as positive and negative controls against the antibody layer (Fig. 4C). As expected, the AuNPs-SAv showed nearly a 200-fold more binding compared to the negative control. Binding in the negative control was attributed to exposed NeutrAvidin and defects, which we identified as uncured PDMS oligomers that leached and settled onto the substrate as aggregates. Extraction of these uncured oligomers via serial solvent exchanges of the PDMS prior to glass bonding reduced the overall incidence of defects (Fig. S3)[40]. As a second validation step, we performed an in-chip dose response assay by assigning each sensing channel to a different concentration in the range of $1.6$-$28.1 \times 10^{10}$ NPs/mL (Fig. 4D). The number of localisations showed a linear dependence up to a concentration of $7.1 \times 10^{10}$ NPs/mL corresponding to 200 counts per 100 $\mu m^2$.

The retrieved particle densities from the dose-response assay defined the upper and lower limits of detection of the platform. Although the optical system had single particle sensitivity, the intrinsic substrate defect density and non-specific binding imposed a lower limit of detection higher than the optical sensitivity, the lowest on the order of 0.5 counts per 100 $\mu m^2$. This problem is common to all label-free approaches based upon elastic scattering. Additional imaging processing can eliminate contributions from intrinsic substrate defects but not from the non-specific bindings. For instance, one could obtain reference image scans of the same area prior to the addition of the analyte of interest and mask out all localisations that were already present in the sample in a routine, analogous to differential-based imaging but with an added step of image registration and alignment. Alternatively, one could switch to a conventional differential imaging approach, i.e., without scanning the FOV across the sample, at the expense of decreasing the throughput.

Regarding the upper detection limit, the likelihood of encountering more than one particle per diffraction limit imposes a boundary to the detectable particle concentration, which for our optical system occurs at values above 200 localisations per 100 $\mu m^2$ (2 per $\mu m^2$). Although by diffraction-limited density considerations alone, the upper value would correspond to about 16 localisations per $\mu m^2$ for a lateral resolution on the order of 250 nm; the experimental value is 8× lower due to a combination of factors: i) the difficulty of packing particles into a dense monolayer due to electrostatic interactions and steric hindrances, ii) the working principle of the single particle localisation algorithm, and iii) the fact that this algorithm operates on single images. Regarding the last two points, the algorithm relies on differentiating between foreground and background pixels to construct a SNR-based threshold – a task that becomes increasingly challenging at higher particle densities. As possible alternatives to extend this upper boundary, one could perform time-lapse differential imaging, akin to super-resolution based imaging; or simply tune the biotin doping ratio (Fig. S4). The latter would allow the sensor to operate in more physiologically conditions for studying protein-protein interactions (i.e. $\mu M$ and mM)[41].

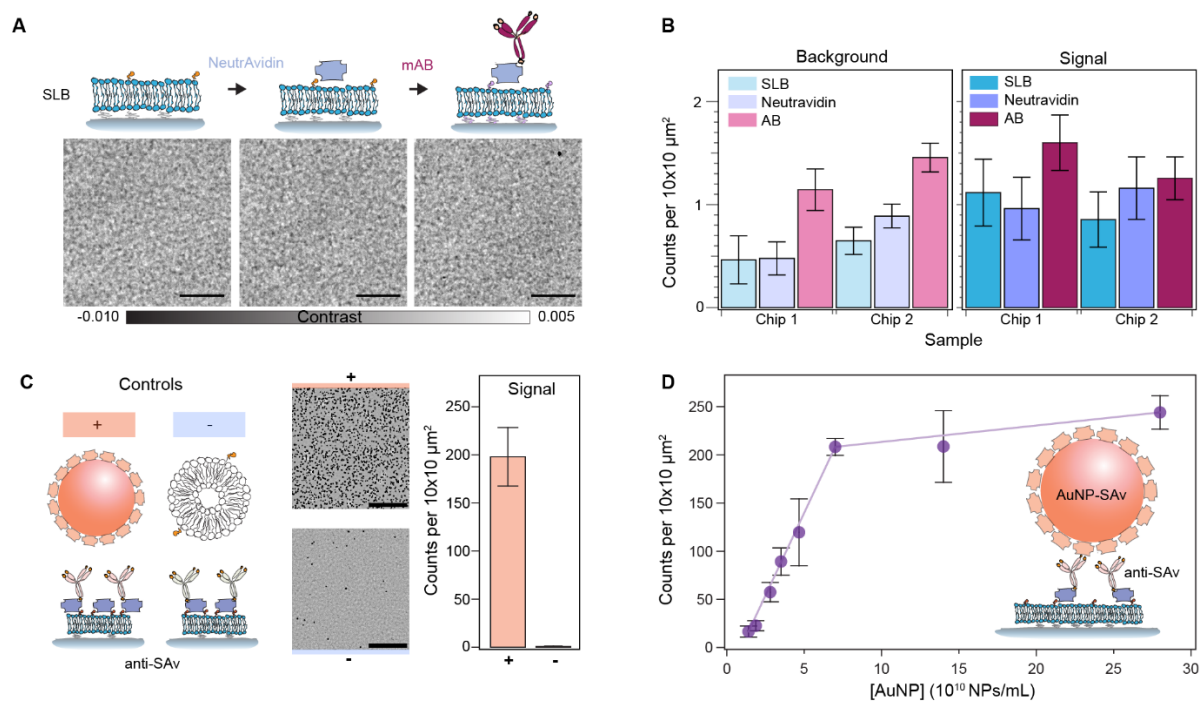

**Fig. 4. On-chip immunoaffinity capture assay validation. (A)** Representative zoom-in images of the substrate after each functionalisation step: bilayer formation, NeutrAvidin incubation, and biotinylated antibody incubation. Scale bars: 5 μm. **(B)** The number of localisations after each functionalisation step. **(C)** Validation of the immunoaffinity functionalisation using streptavidin functionalised AuNPs (AuNP-SAv) and biotinylated liposomes as positive and negative controls, respectively. Scale bars: 10 μm**. (D)** Dose response for different concentrations of streptavidin-functionalised 20 nm gold nanoparticles. Each data point corresponds to the mean of scans covering an area of 0.2 mm$^2$ over multiple different channels within each chip (N = 3). Error bars represent the standard deviation over the mean.

**Immunoassay applied to EV samples**
To demonstrate the compatibility of the sensing platform with complex heterogenous biological nanoparticle systems, we used EVs derived from the ovarian cell line TiOSE4. In detail, we pulled down CD81$^+$ EVs, due to their higher tetraspanin expression levels[42], and studied the resulting binding kinetics as a function of EV concentration and flow rate.

Fig. 5A illustrates an in-chip dose response assay where EVs we continuously introduced at a flow rate of 10.6 μl/h for 8 hours followed by PBS buffer rinsing at a flow rate of 1.3 μL/h for 11 hours. Here the concentration was varied between 1.9×10$^8$ and 7.6×10$^9$ EVs/mL, as determined by NTA, with a different concentration assigned to each sensing channel. In this assay, we followed the surface density of localised CD81$^+$ EVs as a function of time and observed values plateauing between 5 and 9 hours of EV incubation, indicating steady-state conditions. As expected, the associated binding rate and steady-state particle density depended on the EV concentration. Notably, we did not detect significant EV unbinding events during the buffer exchange (Fig. 5A inset) to retrieve a reliable dissociation rate constant, k$_{off}$. Fig. 5B shows the results of three replicates of the in-chip assay together with two additional assays at much higher concentrations to confirm that the same upper limit of detection is reached as in Fig. 4D. The higher degree of variability in the dose response across replicates stemmed from slight chip-to-chip variations, such as effective flow rate. Nevertheless, the intra-chip results show that the sensor can quantitatively determine the relative abundance of EVs expressing a certain target molecule, thus enabling the fingerprinting of EV populations from a panel of surface protein biomarkers.

Overall, the measured binding kinetics are much slower compared to reaction-limited single antibody-antigen interactions and similar surface-based immunoassays[42–44]. We can explain this through a combination of mass transport limited reactions, EV avidity and sensor attinebility. Firstly, the mass-transport limit effectively lowers the association rate, $k_{on}$, expected from reaction-limited kinetics[45]. As a measure of whether our system is reaction or transport limited, we computed the ratio of reaction to mass transport rates characterized by the Dahmköhler number (Da), which was on the order of Da~1-10, with values above one indicating mass transport limited kinetics[46]. In our case the mass transport limited regime stems from the lower analyte flow rates, the high capture probe density of our system, and the lower diffusion coefficient of the EVs relative to typical immunoassays, i.e., 3 $\mu m^2/s$ (EVs) vs 70 $\mu m^2/s$ (proteins). Secondly, the EV avidity reduces $k_{off}$ due to the possibility of multivalent interactions as a single EV can express the same target protein[47]. Thirdly, the high surface coverage of capture antibodies, attinebility, also effectively reduces $k_{off}$ thanks to the increased EV reattachment probability upon unbinding provided by proximal capture antibodies[47].

Together, these three factors lead to complex binding kinetics, which some groups have modelled by introducing additional slow and fast rates for both $k_{on}$ and $k_{off}$[43]. Nevertheless, retrieving reliable rate constants for sensing systems where one of the binding partners is always in great excess relative to the expected dissociation constant, $K_D$, is prone to significant biases. Under these conditions, known as the titration regime, the equilibrium favours the formation of antibody-antigen complexes, and $K_D$ no longer reflects the concentration upon which half the binding sites are occupied[48]. We confirm our system falls within the titration regime by considering that one of the binding partners, the captures antibodies, are approximately three orders of magnitude higher than the typical antibody-antigen $K_D$ values ranging between 0.1-10 nM. We specifically computed the capture antibodies to be in the µM range given an estimated density of 0.08 pmol/cm² and microfluidic channel height of 10 µm. For the sake of simplicity and to focus the scope towards fingerprinting, we restricted the analysis of the binding kinetics to solely determining the time to approximately reach steady state dynamics and the respective binding densities.

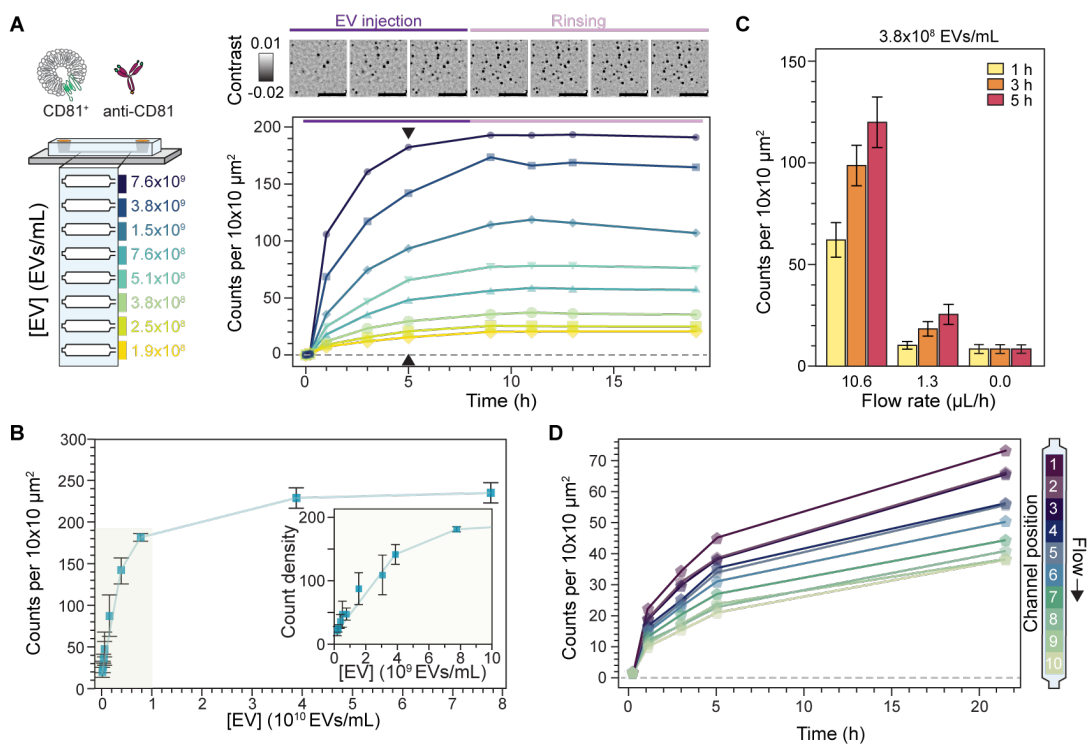

**Fig. 5. In-chip EV binding kinetics. (A)** In-chip dose response assay for CD81$^+$ TiOSE4 EVs with each sensing channel loaded with a different EV concentration. Top: representative zoom-in time-lapse images showing the binding kinetics upon EV injection and subsequent buffer rinsing. Scale bars: 5 μm. Bottom: measured binding kinetics expressed in terms of the number of captured EVs. Each data point corresponds to the mean of scans covering an area of 0.2 mm$^2$. Black arrows indicate the time-point considered as steady-state. **(B)** Corresponding dose response at steady-state. Errors bars indicate standard deviation over the mean (N = 3). **(C)** Effect of flow rate on the binding kinetics at a fixed EV concentration. Each bar corresponds to the mean over 8 independent channel scans, each covering an area of 0.2 mm$^2$. Error bars represent the standard deviation over the mean. **(D)** Spatially resolved intra-channel dose response kinetics under mass transport limited conditions. At flow rates below 1.3 μL/h, EV sample concentration gradients develop as a result of the mass transport limited reaction regime. Each data point corresponds to the mean of a 0.02 mm$^2$ segment of the total scanned area (1/10th) as indicated in the diagram to the left. Arrow indicates the direction of flow, making channel position 1 the entrance of the sensing region.

For mass-transport limited reactions, the flow rates can be further exploited to tune the binding kinetics of the system[49], as shown in Fig. 5C. Namely, increasing the flow rate concomitantly increases the number of captured EVs and pushes the system towards reaction-limited kinetics; albeit at the expense of low sample utilization (capture efficiency). For example, after 5-hour EV incubation, we observed 3-fold (1.3 μL/h flow rate) and 14-fold (10.6 μL/h) improvements of EV capture over no flow conditions. Conversely, decreasing the flow rate exacerbates the mass transport limited binding kinetics thereby increasing the sample utilization (capture efficiency), which in turn leads to analyte concentration gradients along the sensing channel. Nevertheless, because our platform is based upon recording large fields, and keeping this spatial information, these concentration gradients can be exploited for in-channel dose-response experiments as shown in Fig. 5D.

With knowledge of how EV concentration and flow rate affect the binding kinetics, we designed our immunoassay to operate at low flow rates yet with minimal volumes of high sample concentrations. On the one hand, the low flow rates maximise the capture efficiency but have lower overall EV binding densities; while on the other hand, the high EV concentrations (in the range of $10^{10}$ EVs/mL) compensate for the expected lower binding densities, slower kinetics, and allow for low expression biomarkers. That said the platform could have been easily adjusted to target lower EV sample concentrations by increasing the flow rates.

**Molecularly fingerprinting EVs from ovarian cells**

To validate that our optofluidic platform is suited to investigate heterogeneous nanoparticle populations, we tested our system with four different ovarian cell line-derived EVs. Of these ovarian cell lines, three are cancerous (CaOV3, OV90, ES2) and one benign (TiOSE4). In detail, we molecularly profiled these EV populations using a panel of six surface biomarkers and a negative control. We designed a microfluidic chip that integrated all functionalisation steps into a single device, i.e., bilayer formation, immunoassay assembly, and EV immunocapture (Fig. 6A). For molecular profiling, spatially separated sensing channels were independently functionalised with the following antibodies: IgG1 as negative isotype control; anti-CD9, anti-CD63, and anti-CD81, as three classical tetraspanin markers; anti-CD326 (EpCAM), anti-HE4 and anti-CA125, as three ovarian cancer biomarkers. Although not typically associated with EV profiling, both CA125 and HE4 are routinely used in clinical settings to detect ovarian cancer in blood[50].

By imaging areas of 0.2 mm$^2$ per channel, we ensured robust statistics for the fingerprinting assays resulting with an average of more than $10^4$ detected vesicles for each biomarker. Figs. 6B, C show the EV binding kinetics for CaOV3 EVs as a function of positively expressed biomarker together with the total EV counts after 5 hours, here considered as steady state. For the negative control, IgG1, we observed a binding density two- to three-fold higher relative to the baseline defects. This indicated a small degree of non-specific binding of EVs, which is expected upon working at these higher concentrations. Comparison with conventional BSA passivation showed that SLBs minimised non-specific binding on average 4.5-fold better than BSA (Fig. S5). Nevertheless, for all surface biomarkers we detected signals above the negative control. We repeated this measurement with N > 2 replicas for all EV populations and normalised the obtained fingerprint to the average negative control count (Fig. 6D). We opted for such normalisation to determine which biomarkers were positively expressed. For the benign cell line TiOSE4 and the cancer ES2, only the pan-EV tetraspanins were positively detected[51], with the cancer biomarkers showing the same expression levels as the negative control. In contrast, EVs from the cancerous cell lines OV90 and CaOV3 showed positive expression levels for all cancer biomarkers to varying degrees; yet, both EV populations followed a general surface protein expression trend of CD326 >> HE4 > CA125. From a molecular profiling perspective, combining the pan-EV tetraspanins markers with the cancer specific ones, resulted in unique fingerprints for every ovarian cell lines; thus demonstrating the potential of the platform to characterise different subpopulations expressing specific target analytes within a highly heterogenous nanoparticle sample.

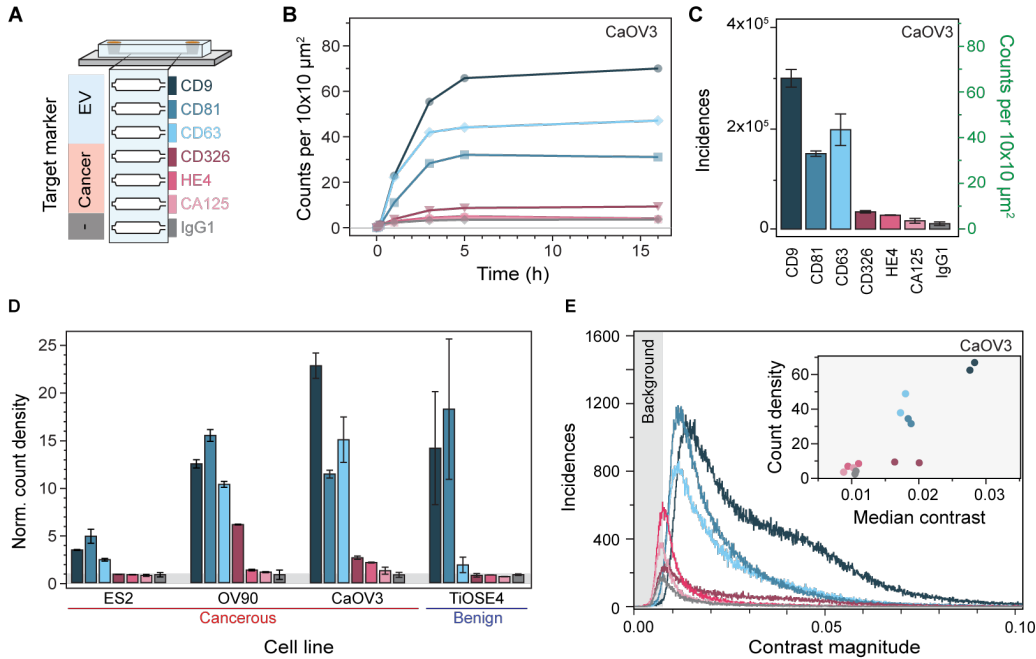

**Fig. 6. Multiplexed EV fingerprinting. (A)** Schematic representation of the multiplexed in-chip immunoaffinity assay used to profile extracellular vesicles expressing different surface markers. Each sensing channel was independently functionalised with a different capture antibody. **(B)** In-chip binding kinetics CaOV3 EVs expressing their respective surface protein markers. **(C)** Molecular fingerprint of CaOV3 EVs expressed in terms of the number of EVs captured within each channel after 5h of continuous flow. The captured vesicles are given in both total and normalised incidences. Errors bars indicate the standard deviation over the mean (N = 3) **(D)** Molecular fingerprint of four different ovarian cell line-derived EVs normalised to the count density of the negative control IgG1. Errors bars indicate the standard deviation over the mean (N = 2). **(E)** Representative contrast distribution for all EVs captured within a single CaOV3 EV fingerprinting assay. Shaded area indicates the background contrast region that is not considered in the EV count density metric. Inset: correlative scatterplot comparing the median contrast magnitude against the count density of each marker expressing EV.

In addition to the total number of EVs expressing a specific surface protein biomarker per channel, our platform also provided insight into the heterogeneity within, and amongst different EV distributions. Namely, Fig.6E shows the distinct contrast distribution from the EVs captured in each channel. To recall, the magnitude of the signal contrast is a function of both the effective refractive index and size of the particles, which makes absolute sizing non-trivial and susceptible with errors. Nevertheless, if we assume that the refractive index of the different sized EVs is similar so that the measured contrast variations are mostly influenced by the EV size, then Fig. 6E inset delivers qualitative information about the relative EV-size differences. For instance, upon plotting the average contrast against the binding densities for the different biomarkers, we observed a positive correlation between larger EVs, associated to higher contrast values, and higher binding densities. We can hypothesise that this correlation is linked to a higher probability of interactions of larger EVs with proximal capture antibodies (attinebility), and possibly also higher number of target surface proteins per EV (avidity).

**Limitations**
In this work, we aimed to develop a platform that reproducibly extracts unique fingerprints with minimal sample volumes from different EV populations based on a panel of surface protein markers. Namely, the platform was optimised to detect surface protein expressions levels of concentrated EV samples ($10^{10}$ EVs/mL) and not to perform absolute EV quantitation of low-concentration samples. Instead, if the goal is either absolute EV quantitation or detection of rare EV biomarkers, then the system should be modified accordingly, by either tuning the microfluidic chip geometry, flow rate, and/or the concentration of capture probes. For instance, to speed up the time response the flow rates can be adjusted to match standard state-of-the-arts which are in the range of 1-10 µL/min. Similarly, the detection of lower concentrations can be accomplished by both increasing the channel height and flow rate to enter the reaction-limited regime.

Regarding the type of sample, the system has been validated with cell line-derived EVs in proof-of-concept experiments. Nevertheless, from a potential diagnostic perspective, the substrate passivation must be improved before working with clinical samples with complex matrices such as blood or plasma. Prior work on non-fluid solid supported lipid bilayers with added PEGylation steps[39] or fast single-step PEGylation with PDMS compatible solvents [52] offer promising routes to improve surface passivation. In addition, the high versatility of the lipid bilayer system offers multiples routes for minimising non-specific binding, and thus improving the overall performance of the platform, such as: i) tuning the lipid composition, ii) modifying the attachment chemistry of the capture probe from neutravidin-biotin linkers to direct covalent interactions, and iii) varying the capture probe from antibodies to aptamers[53,54].

**Conclusions**
This work describes a route for multiplexing and profiling biological nanoparticles in solution based on spatially separated channels on a microfluidic chip. Microfluidic designs presented here could be readily extended with increased the number of independent channels common to more complex microfluidic devices[55], with size and complexity of the device becoming the ultimate limits in terms of multiplexing. Nevertheless, as a complimentary route, our platform is also fully compatible with single-molecule fluorescence read-out approaches[56], and thus could be combined with state-of-the-art fluorescently-tagged antibody[4,57–59] or aptamer[60,61] libraries to enable large scale single particle profiling. We envision that our platform when combined with on-chip standard additions approaches[62] and improved surface passivation could enable diagnostic and care monitoring of diseases based on a selection of disease biomarkers.

In summary, we show that our optofluidic platform integrates necessary assay steps to molecularly profile a population of heterogeneous biological nanoparticles such as EVs in a label-free manner, with single particle sensitivity, robust statistics, and high degree of reproducibility. We demonstrated that our optical read-out allows us to monitor the progress of each step of the assay, and thus optimise the surface functionalisation protocol in terms of robustness, sample preparation time, and high quality of the resulting coatings. We further highlighted the capabilities of our approach to study the underlying heterogeneity of EVs by combining information from the biomarker population as well as its contrast information. We foresee, that upon decoupling the size and refractive index dependence on the particle contrast[25,63,64], these assays would pave the way for a new generation of approaches that can better characterise and study the heterogeneity of EVs by combining molecular fingerprinting, with size and effective material composition information at the single EV level.

**Materials and Methods**

**Reagents.** Bovine serum albumin (BSA, A2934, Sigma-Aldrich) solutions were prepared in phosphate buffered saline (PBS, pH 7.4, 806552, Sigma-Aldrich). Biotinylated neutravidin (10443985, Fisher Scientific) was diluted to a concentration of 0.02 mg/mL (0.3 µM) in 1% BSA. Biotinylated antibodies anti-CD63, anti-CD326(EpCAM), anti-CD9, and anti-CD81 from Ancell (215-030, 126-030, 156-030, 302-030); anti-CA125 and anti-HE4 from LSBio (LS-C86749-1, LS-C743705-50) and Mouse IgG1 k-isotype control from Biolegend (400-104) were prepared to a concentration of 0.05 mg/mL (0.33 µM) in 3% BSA for all experiments. A custom AH peptide with the following sequence: SGSWLRDVWDWICTVLTDFKTWLQSKLDYKD was synthesised by Proteogenix. A stock solution of 1 mg/mL was prepared by dissolving the lyophilised peptide in Milli-Q water according to manufacturer's recommendation. This stock solution was aliquoted and stored at -20°C for up to one month. For all experiments, the peptide stock solution was diluted to 0.45 mg/mL (200µM).

**Liposome preparation.** All liposomes were composed of 1-palmitoyl-2-oleoyl-sn-glycero-3-phosphocholine (16:0-18:1 PC) (850457C, Avanti Polar Lipids, Inc) doped with 1,2-dioleoyl-sn-glycero-3-phosphoethanolamine-N-(cap biotinyl) (18:1 Biotinyl Cap PE) (870273C, Avanti Polar Lipids, Inc) in a 99:1 molar ratio. To prepare the liposomes, the two lipid stock solutions in chloroform were first mixed and subsequently dried with a nitrogen stream and then placed under vacuum for 24 hours. The dried lipids were then rehydrated to a concentration of 5 mg/mL in TRIS Buffer (100 mM, pH 7.4, 648315, Sigma Aldrich) and vortexed for 2 minutes. These solutions were stored in the freezer for up to 1 month. Liposomes were then prepared using two approaches: sonication and extrusion. For bathsonication, the 5 mg/mL lipid solution was sonicated for 20 minutes at room temperature. For extrusion, the hydrated lipid solutions were passed 21 times through polycarbonate membranes, ranging in size from 200 nm to 30 nm, using Avanti Polar Lipids Mini-Extruder (610000). To prepare the smaller liposomes, i.e., 50 nm and 30 nm, the vesicle suspensions were serially extruded through successively smaller membrane pore sizes. All experiments were performed using vesicle suspensions at 1 mg/mL. Finally, 5 µL of a 500 mM $CaCl_2$ solution in Milli-Q water (Millipore) was added to the 500 µL 1 mg/ml liposome dilution. Unless stated otherwise, all liposome solutions were used within 5 days of preparation to minimise ageing effects.

**Fabrication of microfluidic chips.** Microfluidic chips (MF) were fabricated using two-layer soft lithography. Two molds were made on silicon wafers using a laser writer (Heidelberg uMLA, 365 nm), one for the flow layer using AZ P4620 (Microchemicals, GmbH) and one for the control layer using SU8 1060 (Gersteltec). The MF chips are made from polydimethylsiloxane (PDMS, Sylgard 184) mixed at a ratio of 10:1 polymer to curing agent. To make the thinner flow layer, the PDMS was spin coated onto the wafer resulting in a thickness of about 30 µm. For the thicker control layer, the PDMS was dropcast onto the control mold to achieve a thickness of about 5 mm. The PDMS was then degassed under vacuum for two hours before baking in a convection oven at 80°C for one hour. Once cured, the PDMS on the control mold was pealed from the wafer and the resulting control chips were cut out and holes were punched. The control layer chips together with PDMS covered flow wafer were treated with oxygen plasma (Diener Electronic, Atto 13.56 MHz, 10.5 L) for 1 minute (300 W, 1.5 sccm) before being aligned under a stereo microscope. To bind them together, the aligned chips were baked for one hour in an 80°C oven. The bound chips were removed from the flow wafer and the holes in the flow layer were punched. To complete MF chip assembly, the resulting two-layer MF chips and cleaned glass coverslips (24 × 40 mm$^2$, 0.17 mm, Karl Hecht) were exposed to oxygen plasma for 1 minute, bound together, and baked at 80°C for one hour.

**Fabrication of microwells.** The microwells were made by punching 5 mm holes into unpatterned cured PDMS of the same thickness. These were bound to cleaned glass coverslips using the same process described for the MF chips.

**On-chip bilayer formation and immunoassay functionalisation.** To begin the on-chip bilayer formation, reagents were loaded into medical grade microfluidic tubing (AAD04103, Tygon) and connected to the MF chip. For the peptide specifically, the tubing was first primed with 3% BSA solution for 30 minutes to reduce non-specific binding. To begin, the chip was primed with PBS until all the air within the channels was removed. Then the liposomes were flowed into the channels until the bilayer had visibly formed, approximately 1 minute under our experimental conditions. Once bilayer formation had occurred, the channels were rinsed with PBS for 1 minute to remove any excess unbound liposomes. After rinsing, the AH peptide was flowed through the channel for approximately 1 minute, or until no further vesicle rupture was visible. The bilayer was then rinsed with PBS for 5 minutes to remove all the peptide. This results in a fully formed bilayer coating. Next NeutrAvidin was flowed into the channel for 3 minutes and incubated for 30 minutes. The channel was again rinsed with PBS for 5 minutes and then the chosen capture antibodies were flowed into the channels for 3 mins and incubated for 30 minutes. A final rinsing step with PBS for another 5 minutes completed the immunoassay. For the fingerprinting assay the target EV sample was flowed at a rate of 1.3 µL/h for at least 5 hours. Specifically, the EV stock solutions were diluted in 0.1% BSA to a target concentration in the range of 2-4×10$^{10}$ EVs/mL on the day of experiment.

**Microwell bilayer formation.** To begin the bilayer formation, the microwells were primed with 20 µL of PBS. Then 20 µL of 1 mg/mL of liposomes were injected. Once bilayer formation had occurred, 20 µL of liquid was removed prior to washing. Washing involved adding 50 µL of PBS into the microwell, followed by pipetting up and down 10 times and subsequently removing 50 µL of liquid. This process was repeated five times. Then 20 µL of 200 µM AH was added and mixed by pipetting up and down 10 times. The peptide was left in the microwell between 10 seconds to 1 minute depending on its efficiency in rupturing liposomes. Finally, an additional step of washing was performed, resulting in the final bilayer.

**EV isolation from cell lines.** The human ovarian cancer cell lines, including CaOV3, OV90, and ES2, were purchased from American Type Culture Collection (ATCC). The benign cell line, TiOSE4, was obtained from transfection of hTERT into NOSE cells maintained in 1:1 Media 199: MCDB 105 with gentamicin (25 µg/mL), 15% heat-inactivated serum, and G418 (500 µg/mL) (Clin. Cancer Res. 2015, 21, 4811−4818). CaOV3 and ES2 cell lines were cultured in DMEM (Hyclone) and McCoy's 5A (Gibco), respectively. In addition, OV90 and TiOSE4 cell lines were maintained in RPMI-1640 (Hyclone). All basal media were supplemented with 10% fetal bovine serum (FBS, ThermoFisher Scientific), 100 U/mL penicillin, and 100 μg/mL streptomycin (Cellgro) at 37°C in 5% $CO_2$. EVs were isolated as previously reported[65]. In brief, cells were cultured to 80-90% confluence in a basal medium and washed with PBS to remove unattached cells and debris. Next, the cells were incubated in a conditioned medium supplemented with 1% Exosome-depleted FBS (ThermoFisher Scientific), 100 U/mL penicillin, and 100 μg/mL streptomycin for 48 hours. The medium was collected and centrifuged with 300 × g for 5 minutes at 4°C to remove floating cells or large debris. The supernatant was passed through a 0.8 μm membrane filter (Millipore Sigma) and concentrated using a Centricon Plus-70 centrifugal filter (MWCO = 10 kDa, Millipore Sigma) with 3,500 × g for 30 minutes at 4°C. The sample was then loaded onto the size-exclusion chromatography (SEC) column packed with 10 mL of CL-4B Sepharose (Cytiva). The fractions of 4 and 5 (a total of 2 mL) were collected and concentrated with the Amicon Ultra-2 Centrifugal Filter (MWCO = 10 kDa, Millipore Sigma). The 1x protease and phosphatase inhibitor was added and stored at -80°C until use.

**EV characterisation.** EVs were lysed in LIPA lysis buffer (Cell Signaling Technology) for western blot analysis to confirm the characteristic EV biomarkers (CD9, CD63, and CD81). The blots were probed with flowing primary antibodies: anti-CD9 (1:500 dilution, BD Biosciences), anti-CD63 (1:500 dilution, Ancell), and anti-CD81 (1:500 dilution, Santa Cruz Biotechnology). Chemiluminescence was detected using an Azure 280 imaging system (Azure Biosystems). The concentrations and sizes of EVs were determined by nanoparticle tracking analysis using Nanosight NS300 and were found to be $1.22 \times 10^{11}$ particles/mL (CaOV3), $2.7 \times 10^{11}$ particles/mL (OV90), $3.0 \times 10^{11}$ particles/mL (ES2), and $3.8 \times 10^{11}$ particles/mL (TiOSE4)

**Microscope.** The custom-built spatially incoherent digital holographic optical system was based on a common-path microscope operating in reflection, whereby illumination and imaging arms were separated by a single 50:50 beamsplitter plate (BSW27, Thorlabs) and all optics were arranged in a 4f configuration. In the illumination arm, a 455 nm light emitting diode (M455F3 LED, Thorlabs) was coupled into a 200 μm multi-mode fibre (M25L02, Thorlabs). Light outcoupled from the fibre using a 6.24 mm aspheric lens (A110TM-A, Thorlabs) was then relay imaged onto the sample plane formed by a 1.46 NA oil immersion objective (APON 60XOTRIF, Olympus) via a 1:1 imaging system, composed of two 300 mm achromatic doublet lenses (AC508-300A, Thorlabs). Under this optical arrangement the NA of illumination was approximately 0.5, resulting in a flat-top illumination with a diameter of 89.5 μm. For the imaging arm, light collected from the sample by the objective and reflected off the 50:50 beamsplitter was imaged onto a scientific CMOS camera (C11440-22CU, 6.5 μm pixels, Hamamatsu) using a 300 mm achromatic doublet (AC508-300A, Thorlabs) resulting in a 100× magnification. The sample was mounted on a motorised XY microstage (Mad City Labs) equipped with linear encoders, as well as a XYZ nanostage (Nano-LP200, Mad City Labs). The sample focus position was stabilised to within 10 nm using the backreflection from 670 nm misaligned confocal beam with a low numerical aperture of illumination (CPS670F, Thorlabs). Specifically, the beam position was used as a feedback parameter in the proportional–integral–derivative loop.

**Optical imaging.** For all experiments, we measured a power at the sample between 1.4-1.7 mW equivalent to an irradiance of 0.22-0.27 μW/μm². During acquisition, a field of view of 66.6 μm × 66.6 μm corresponding to an area of 1024×1024 camera pixels was recorded with an exposure time of 10 ms and a fixed frame rate of 100 Hz. To minimise data load and increase the signal to noise ratio, the data were saved in the form of 100 time-averaged frames, leading to an effective time resolution of 1 Hz. Prior to each data acquisition, an experimental flat-field image was generated and saved. The flat field imaged was produced by first collecting a stack of at least 60 time-averaged frames taken at different sample locations and same focus position, and subsequently taking the median value on a pixel-by-pixel value. This flat field image contained inhomogeneities along the optical system and imperfections in the sample illumination.

**Image processing.** All images were first normalised to the average background camera counts in the background. Next, we flat field corrected the normalised images, by division, to remove inhomogeneities attributed to the optical system and sample illumination. For image scans, the image stacks were stitched together using a phase correlation algorithm. To remove large feature contributions from the flat field corrected images, such as out-of-focus objects corresponding to the top surface of the PDMS microfluidic device, a spatial median filter with a kernel size of 17 pixels was determined and subtracted. This process had no effect on contrast or shape of the diffraction limited spots.

**Particle localisation.** First a global noise level from each image was estimated from the median absolute deviation. Next, a local noise estimate of each image was determined by computing the root-mean-square of all pixel values within a kernel size of 65 pixels falling within 2.5× the global noise estimate. This local noise estimate was then used to determine a signal to noise ratio image; i.e, by dividing the initial processed image by the estimated local noise. Then, candidate regions of interest were segmented based on the following two selection criteria: a) pixel based: positive for all pixels exceeding a signal to noise threshold of 4; and 2) clustering based: positive if there were a minimum of 3 pixels exceeding the SNR threshold within a 3 × 3 pixel$^2$ area. Diffraction limited spots satisfying the selection criteria were then segmented and localised with sub-pixel precision using the radial symmetry centres algorithm[66]. The resulting lateral position, signal contrast and the integrated signal contrast were stored for further processing.

**Supporting Information**.

The Supporting Information is available free of charge.

- Equivalence between spatially incoherence illumination schemes (Figure S1), effect of liposome ageing on SLB formation (Figure S2), effect of different PDMS cleaning strategies on substrate cleanliness (Figure S3), lipid composition of the SLB determines the capture site density (Figure S4), non-specific binding comparison between SLB and BSA (Figure S5), and respective standard EV characterisation (Figure S6). (PDF)
- Captions for Movies S1 to S2

**Author Contributions**

Conceptualization: HI, HL, JOA, RQ
Methodology: AS, JGG, JOA

Investigation: AS, JSH, HI, JOA
Software: JOA
Formal analysis: AS, JSH, JOA
Visualization: AS, JOA
Supervision: HI, HL, JOA, RQ
Funding acquisition: HI, HL, JOA, RQ
Writing—original draft: JOA
Writing—review & editing: AS, JGG, JSH, HI, HL, JOA, RQ

**Funding Sources**

Swiss National Science Foundation grant 207485 (JOA, RQ)
US National Institutes of Health grant R01CA229777 (HL)
US National Institutes of Health grant U01CA233360 (HL)
US National Institutes of Health grant R01CA239078 (HL)
US National Institutes of Health grant R01CA237500 (HL)
US National Institutes of Health grant R01CA264363 (HL)
US National Institutes of Health grant R21CA267222 (HL)
US National Institutes of Health grant R21CA217662 (HI)
US National Institutes of Health grant R01GM138778 (HI)

**Competing interests:** Authors declare that they have no competing interests.

**Data and materials availability:** All data needed to evaluate the conclusions in the study are present in the paper and/or the Supplementary Materials.